\documentclass[11pt,a4paper]{article}

\usepackage[T1]{fontenc}
\usepackage[utf8]{inputenc}
\usepackage{lmodern}
\usepackage[a4paper,margin=2.5cm]{geometry}
\usepackage{setspace}
\usepackage{booktabs}
\usepackage{array}
\usepackage[round,authoryear]{natbib}
\usepackage{authblk}
\usepackage{xcolor}
\usepackage[hidelinks]{hyperref}
\usepackage{microtype}
\usepackage{float}

\providecommand{\Description}[1]{}

\title{\textbf{Beyond the principle--practice gap: Responsible AI as situated epistemic and ethical practice}}

\author[1]{Thao Do-Ngoc\thanks{Corresponding author: \href{mailto:tnd32@bath.ac.uk}{tnd32@bath.ac.uk}. ORCID: \href{https://orcid.org/0000-0002-0015-892X}{0000-0002-0015-892X}}}
\author[1]{Emma Carmel\thanks{\href{mailto:E.K.Carmel@bath.ac.uk}{E.K.Carmel@bath.ac.uk}. ORCID: \href{https://orcid.org/0000-0002-3170-3150}{0000-0002-3170-3150}}}
\affil[1]{University of Bath, Bath, United Kingdom}
\date{Preprint, \today}

\begin{document}
\maketitle

\begin{abstract}
\noindent Responsible AI scholarship has found that ethical principles are frequently not reflected in AI design and development. Conceptualised as a `gap' between principle and practice, this is framed as a problem of implementation, awareness or compliance, and by implication, a deficit of systems or individual behaviour. This paper sets out an alternative explanation. Drawing on fourteen semi-structured interviews with actors designing and governing AI-enabled anti-trafficking systems across law enforcement, commercial, research, and survivor-support contexts, it argues that such apparent gaps emerge from the varied and complex institutional and professional conditions through which actors come to recognise some issues, but not others, as ethical problems requiring action. We show that epistemic choices are intertwined with ethical judgements, and trace four interconnected moments---perception, judgement, situated action and socio-technical materialisation---through which concerns become intelligible, are weighed and acted upon, and are embedded in technical artefacts and governance arrangements. Actors often hold genuine ethical commitments, but these commitments are shaped by institutional logics and professional moral economies: some concerns are materialised, while others are narrowed, deferred or left unrealised. We show that ethical agency is exercised within and between competing constraints of profession, role, prior experience, and institutional settings. The paper argues that responsible AI governance requires purposeful and well-developed epistemic infrastructures that can cut across these constraints to establish institutional conditions through which the knowledge of affected communities and potential harms becomes receivable and actionable in design.

\medskip
\noindent\textbf{Keywords:} responsible AI; institutional logics; moral economy; situated agency; anti-trafficking
\end{abstract}

\section{Introduction}

Responsible AI scholarship has moved beyond articulating high-level principles, such as fairness, accountability, transparency, privacy and human oversight, towards asking how these principles are understood and enacted in practice. A substantial body of work documents persistent difficulties in operationalising them, including the abstraction of principles, conflicts among values, the lack of appropriate metrics, fragmentation across development pipelines, and weak organisational advocacy and accountability \citep{sadek2024challenges}. Some locate the problem in practitioners' ethical capacities and call for stronger ethical awareness, professional virtues and practical wisdom \citep{hagendorff2020ethics,hayes2024applied}. Others emphasise the organisational and structural level, notably in critiques of ``ethics washing'', where ethical commitments are adopted without altering underlying priorities or practices \citep{biddle2024can,bietti2020ethics,schultz2024digital,maanen2022ai}. From this perspective, responsible AI requires stronger institutional and governance arrangements rather than reliance on individual ethical commitment alone \citep{maclure2025ai,mittelstadt2019principles}.

Empirical work has further shown that responsible AI is organisationally and politically situated. Ethical initiatives operate within existing structures and incentives \citep{rakova2021where}; fairness checklists function as organisational infrastructures rather than self-sufficient technical mechanisms \citep{madaio2020co}; user experience (UX) practitioners engage in forms of ``soft resistance'' within organisational constraints \citep{wong2021tactics}; ethics practitioners act as ``ethics entrepreneurs'', often only partially connected to core organisational decisions \citep{ali2023walking}; and even the ``license to critique'' is unevenly distributed within development teams \citep{widder2024power}.

Yet three related issues remain insufficiently resolved. \emph{First,} framing the challenge as a ``gap'' can obscure how understandings of (ethical) responsibility themselves develop through practice. Principles do not exist as stable commitments upstream of development, which are then implemented more or less successfully downstream. Rather, principles and practices are recursive, as ethical commitments are clarified, contested and reformulated when actors encounter particular technologies, datasets, users, and practical problems \citep{moats2025valuable,ruster2025gaps}. The question is therefore not only whether principles reach practice, but how understandings of responsible action are formed and transformed through practice. \emph{Second,} much of this constitution happens through choices that are not usually treated as ethical. Throughout AI development, developers have to decide which problems to make computationally tractable, which data and proxies can stand for them, what counts as ground truth and what a system may legitimately infer. These are epistemic choices about what can be known and represented. Yet they are often approached as technical matters, while ethics is assessed separately or afterwards against the resulting system. Scholars have argued that the epistemology and ethics of AI cannot be separated \citep{pozzi2024ethics,russo2023connecting}. What remains less understood is how practitioners themselves recognise and negotiate this entanglement, and what happens to an ethical concern that arises in the form of an epistemic one. \emph{Third}, empirical research has concentrated on commercial technology organisations, where divergence between commitment and practice can plausibly be explained by commercial incentives, suppressed voice or weak authority. This leaves a different puzzle: how should variation in responsible practice be explained when actors are strongly committed to a social or ethical purpose that is central to their technologies or organisations? Shared commitment does not necessarily generate a shared understanding of what responsible action requires. At the same time, research increasingly recognises that institutional contexts shape how actors interpret principles and translate them into governance instruments \citep{bargil2025examining,minkkinen2023institutional,palladino2023biased}. Multiple institutional logics may coexist within an organisation and similar logics may operate across organisations \citep{besharov2014multiple,osce2020leveraging}. Less attention has been paid to how actors working on the same social problem, but situated within different institutional contexts, arrive at different epistemic and ethical judgements. This is particularly relevant as AI systems are increasingly developed and used across institutionally plural domains, including education, social protection and criminal justice. Understanding responsible AI in these settings requires moving beyond the organisation as the unit of analysis to the institutional and professional contexts through which actors interpret problems, evaluate responsibilities and negotiate their actions.

Drawing on 14 semi-structured interviews with actors involved in the design, development, evaluation, and governance of AI-enabled anti-trafficking systems\footnote{The term \emph{AI-enabled anti-trafficking systems} refers to socio-technical systems that incorporate artificial intelligence techniques, such as machine learning, natural language processing, computer vision or predictive analytics, to support efforts to prevent, identify, investigate or respond to human trafficking.}, this paper asks: \emph{How do actors negotiate epistemic and ethical judgements and their translation into socio-technical design and governance practices?}

Anti-trafficking provides a revealing setting to examine these issues. Rather than being organised around a single profession, technology or model of intervention, the field is institutionally plural, bringing together law enforcement and intelligence actors; civil society and survivor-support organisations; supply chain and compliance actors; researchers and technology developers, often through cross-sector collaboration \citep{bryant2020combatting,foot2015developments,osce2020leveraging}. While these actors share a broad commitment to addressing exploitation, they operate within different institutional and professional worlds that shape how evidence, harm, vulnerability, responsibility and legitimate intervention are understood. Critical scholarship has long documented tensions around trafficking, migration, sex work, policing, victimhood and protection \citep{akhmetova2021chapter,davis2024prosecution,demeke2024human,kjellgren2022good,mai2022migration}. Responsible AI principles therefore enter a field characterised not by an absence of ethical concern, but by competing understandings of what ethical action requires and how it should be prioritised. These challenges are further complicated by the domain's distinctive epistemic difficulties. Human trafficking is difficult to observe and measure directly, its legal and empirical boundaries are contested, and affected populations are often hidden or difficult to identify \citep{cockbain2022examining,gallagher2012international,schroeder2022review}. Official datasets capture identified or assisted victims rather than the population experiencing trafficking, and identification and reporting practices shape what becomes visible in the data \citep{altammemi2023seeing}. Digital traces, such as online advertisements and platform activity offer new sources of information, but interpreting them as indicators of trafficking remains epistemically and politically contested. For example, features of escort advertisements operationalised as trafficking indicators also occur in consensual sex work \citep{giommoni2024cannot,lugograulich2024improving}, while commonly proposed labour-trafficking indicators appear in ordinary migrant job advertisements \citep{volodko2019spotting}. Such ambiguities cannot be resolved through technical performance alone, particularly when the use of these indicators may support the identification of exploitation while also creating risks of surveillance, profiling and misclassification \citep{deebswihart2022ethical,kjellgren2022good,musto2019limits}.

It is through these intersections of epistemic uncertainty, ethical commitments and institutional plurality that this paper examines how responsible AI practices take shape. We trace four analytically distinguishable but recursive moments. \emph{Perception} concerns what becomes intelligible as an epistemic or ethical issue; \emph{judgement}, how concerns acquire significance and priority; situated \emph{action,} actors' capacity to negotiate and respond; and \emph{socio-technical} \emph{materialisation}, whether judgements become embedded, narrowed, deferred, delegated or rejected in technical and governance arrangements.

\section{Conceptual foundation}

We conceptualise responsible AI as an institutionally situated practice that is both epistemic and ethical. Rather than treating it as the implementation of general principles, we focus on how concerns become visible, are judged significant, are made actionable, and are inscribed into socio-technical systems. Values-in-design and socio-technical scholarship has shown that values are raised, negotiated and materialised through design practices, technical choices and organisational arrangements \citep{akrich1992de,shilton2013values,suchman2007human}. In this process, developers face consequential decisions about problem definitions, relevant knowledge and data, categories and proxies, and the inferences and uses their systems are intended to support. We, therefore, distinguish analytically between epistemic considerations---what can legitimately be known, represented or inferred through data and computation, such as the adequacy of categories, proxies, ground truth and evidence---and ethical considerations---the harms, rights, obligations and responsibilities associated with those choices and the system's anticipated uses.

As \citet{mittelstadt2019principles} argues, the practical meaning of responsible AI principles must be interpreted through context-specific norms. To examine how these processes unfold within particular professional and institutional contexts, we draw on institutional logics, moral economy, sensemaking and situated agency. As AI technologies enter professional fields, they arrive as unsettled socio-technical objects. By \emph{unsettled}, we mean that their meaning, evidentiary status and appropriate use are not yet stabilised in professional routines. Established registers have ways of handling familiar objects---a witness statement, a disclosure or a supplier risk---but risk scores, anomaly signals, scraped images, or vulnerability indicators fit these practices more or less easily. The epistemic--ethical distinction is particularly salient in socially sensitive domains such as anti-trafficking, where the phenomena being represented are contested, difficult to observe and closely connected to questions of vulnerability, rights and intervention \citep{cockbain2022examining,demeke2024human,mai2022migration}.

\subsection{Institutional mediation of epistemic and ethical considerations}

Moral economy directs attention to the values, obligations, emotions, and professional commitments through which communities distinguish good, responsible, or legitimate activity from what is unacceptable, careless, or harmful \citep{daston1995moral,fassin2020two}. \citet{daston1995moral} defines it as a collective \emph{``web of affect-saturated values''} regulating what counts as good work while \citeauthor{fassin2012humanitarian}'s \citep{fassin2012humanitarian} account of humanitarian reason extends the concept to the governance of vulnerability, showing how compassion toward suffering can simultaneously legitimise both practices of care and control. Anti-trafficking is a clear case where intervention framed through rescue and protection may be channelled into surveillance, policing or restrictive governance \citep{deebswihart2022ethical,mai2022migration}. Critical scholarship has shown that anti-trafficking governance has often made some harms and subjects more intelligible---and therefore actionable---than others. Drawing on \citeauthor{christie1986ideal}'s \citep{christie1986ideal} ``ideal victim'' and its later extension \citep{duggan2018revisiting}, scholars have shown how anti-trafficking measures have privileged the innocent, feminised, sexually exploited victim, while labour exploitation, migrant precarity, sex worker safety and male victimhood remain less visible \citep{forringerbeal2022ideal}. \citet{islam2026beyond} show how a ``moral economy of victimhood'' translates judgements of innocence and credibility into institutional recognition. Such hierarchies of victimhood, rescue, protection, criminality and deservingness form the evaluative terrain within which concerns about AI become perceptible and acquire normative weight.

These orientations are socially and institutionally situated. Hoffman defines institutions as \emph{``rules, norms, and beliefs that describe reality for the organization''} \citep[p.~351]{hoffman1999institutional}, while institutions can guide behaviour without requiring \emph{``repeated authoritative intervention''} \citep[p.~445]{clemens1999politics}. Institutional logics---the material practices and symbolic systems through which communities give meaning to activity \citep{thornton2012institutional}---provide socially available repertoires for interpreting situations.

An investigative logic may foreground detection, evidence, and prosecution; a protection logic may prioritise safety, care, and harm avoidance; commercial and compliance logics may emphasise risk management and auditability. Logics thus make certain ways of seeing and acting appear natural, legitimate, or necessary. Actors may draw on multiple, sometimes competing, logics when making sense of the same situation. Even so, AI inputs and outputs such as risk scores, classifications or inferred indicators do not carry self-evident meanings. Sensemaking describes how actors extract cues, construct plausible understandings of unfamiliar outputs and decide what should follow \citep{weick1995sensemaking,weick2004sensemaking}. We treat sensemaking not as neutral individual cognition but as drawing on the moral economies and logics in which it is embedded.

\subsection{Situated action and socio-technical inscription}

\citet{suchman2007human} challenges the view of action as the execution of predefined plans. Rather, action is accomplished through ongoing engagement with particular social and material circumstances, and organisational activities emerge through the interrelations of human actors, technologies and material arrangements \citep{orlikowski2007sociomaterial}. In AI development, developers might encounter data limitations, technical constraints, or competing organisational priorities that can alter both what actions seem possible and what responsible practice means \citep{madaio2020co,moats2025valuable}. Yet a concern about privacy or harm may shape a conversation without altering the arrangements through which a system operates. We therefore draw on \emph{inscription:} designers inscribe into artefacts a ``script'' of assumptions about the world, users, and the actions the artefacts afford or foreclose \citep{akrich1992de}, and design practices can act as ``values levers'' that turn ethical reflection into features and decisions \citep{shilton2013values,winner1980do}. Along the way, concerns may be strengthened, narrowed, deferred, redirected or lost. We ask which concerns become stabilised (or not stabilised) in socio-technical arrangements, in what form, and with what degree of durability.

\subsection{Reflexivity and embedded agency}

Actors are neither passive carriers of institutional expectations nor fully autonomous ethical decision-makers. Institutional scholarship addresses this as the problem of embedded agency: how actors shaped by institutions can reflect on, maintain or challenge them \citep{battilana2009institutional,hargrave2009institutional}. \citet{battilana2009institutional} highlight how institutional contradictions, exposure to multiple logics and actors' social positions can create opportunities for agency, and reflexive deliberation mediates between structural conditions and conduct, allowing actors to evaluate their circumstances \citep{archer2003structure}. In AI development, this may involve questioning problem definitions, the adequacy of data or tensions between institutional priorities and the consequences of system design. However, agency here should not be equated with resistance or improvement: actors may actively reproduce institutional arrangements even while recognising their limitations. We therefore ask under what conditions actors exercise situated agency in responding to epistemic and ethical concerns.

\section{Methodology}
\subsection{Case selection}

Since the mid-2010s, there has been growing interest in the use of AI and data-driven technologies in anti-trafficking work, spanning online investigation, image and text analysis, supply-chain risk detection, victim identification, chatbot-based victim support, worker voice, case management and data coordination \citep{osce2020leveraging}, across the 4Ps of prosecution, prevention, protection and partnership \citep{foot2015developments}. Stated benefits include identifying patterns in large-scale online data, reducing investigators' exposure to harmful content, supporting supply chain risk assessment, and widening access to support \citep{pradhan2026online,wang2026ai}. However, anti-trafficking interventions have long been criticised for legitimising surveillance, policing and migration enforcement, sometimes harming marginalised populations, including migrants, sex workers and racialised communities \citep{chuang2014exploitation,mai2022migration}. AI may amplify these tensions while intensifying risks of misclassification, profiling, decontextualisation and weakened accountability \citep{akhmetova2021chapter,deebswihart2022ethical,kjellgren2022good,nair2025ask}. We approach anti-trafficking as a distributed institutional field, rather than a single coherent domain, in which actors operate under different mandates, professional norms and institutional logics. Multiple understandings of trafficking, vulnerability, victimhood, evidence and intervention can coexist and compete.

A recent field-level analysis of AI-enabled anti-trafficking systems identified the coexistence of multiple institutional logics within the field \citep{dongoc2026selective}. These include prosecution logics that frame trafficking as crime requiring evidence; humanitarian logics that frame it as harm requiring protection; compliance logics that frame it as liability requiring management; and data infrastructure logics that frame it as information coordination requiring integration. Each logic shapes which problems are formulated, which data are treated as relevant, how labels are authorised, and how outputs are interpreted. The present study shifts the analytical focus from field-level patterns to situated actors, examining how institutional logics and moral economies are interpreted, negotiated, enacted and sometimes challenged through practitioners' situated epistemic and ethical choices.

\subsection{Research design and sampling}

This study adopts an interpretive design and qualitative research methods to examine how actors involved in AI-enabled anti-trafficking systems perceive, judge and act upon epistemic and ethical issues. We conducted 14 semi-structured interviews between October 2025 and February 2026, lasting 30-90 minutes, recorded and transcribed with consent. Participants were involved in designing and developing AI-enabled anti-trafficking systems, alongside actors with experience in research, deployment and governance. Individual interviews were chosen to elicit detailed accounts of how participants defined problems, evaluated evidence, negotiated competing priorities and justified development decisions. The semi-structured format allowed an open, exploratory approach, and individual conversations enabled us to probe reasoning and experiences that might be difficult to discuss in group settings.

The sampling frame was designed to maximise analytically relevant heterogeneity. To capture the field-wide diversity identified above, we sought participants from different organisations involved in different forms of intervention and computational practice: investigative and intelligence applications; supply chain and ESG risk assessment; survivor and worker support technologies; research and data governance and field-level partnership. Participants' roles, expertise and decision-making responsibilities were a secondary criterion; the importance of professional biographies emerged in the analysis as discussed below. Demographic characteristics were not used as sampling criteria. This heterogeneity enabled comparison of how epistemic and ethical questions arose across cases, without treating individual participants as representative of wider professional or organisational groups.

Recruitment combined purposive and snowball sampling: six participants were approached through professional connections, three were identified and approached through independent desk research, and five through referrals from interviewees. The sample included technology developers, researchers, civil society actors, law enforcement officers, and commercial actors (Appendix~\ref{app:participants}). Projects ranged from concepts and prototypes to piloted, scaled, concluded or discontinued systems. Some participants discussed several projects, broadening the range of development decisions and organisational circumstances. As the field is small, participants are identified by letter codes and organisational and project details are reported at a level that limits re-identification. The study received institutional ethics approval and safeguarding information was made available.

\subsection{Data analysis}

Analysis followed an iterative and abductive approach. Initial coding was specific and stayed close to the interview material, identifying how participants described their systems, problems, data and the issues they encountered, with attention to moments in which they justified choices about problem definitions, categories, proxies, uncertainty, harms, responsibility and use. Subsequent analysis traced how choices about what could be represented or inferred connected to questions of harm, rights, and legitimate intervention, and which professional and institutional contexts were invoked. Moral economy, institutional logics, sensemaking and situated agency were used as sensitising concepts. Material that complicated or contradicted emerging interpretations was explored in depth to highlight areas of tension and points of divergence. Cross-case comparison also grouped recurring orientations into four moral-economy registers according to the forms of harm, responsibility and legitimate knowledge participants foregrounded. Comparison across cases produced the four-moment framework that organises the findings.

\subsection{Limitations}

The study relies on stakeholder accounts rather than observation of system development, deployment or use, so it examines participants' interpretations and justifications rather than assessing system performance or downstream impacts. It does not include survivors, workers or other affected communities; therefore, the analysis cannot capture their lived experiences. However, it is designed to examine how their perspectives are recognised, represented or overlooked by institutional actors. The relatively small and heterogeneous sample supports theoretical development by examining variation across professional and institutional settings rather than generalisation.

\section{Findings: the entanglement of epistemic and ethical considerations}

Across the interviews, epistemic and ethical considerations emerged as closely intertwined. How trafficking was defined, represented and inferred shaped what participants understood as harms and legitimate interventions; conversely, ethical commitments influenced what participants considered appropriate to know, which data should be collected, connected or withheld, and what forms of computational inference could be justified. Concerns about privacy, participation, autonomy and protection often generated epistemic responses: revising data collection procedures, seeking alternative forms of domain knowledge, restricting data linkage or reconsidering what information a system needed to produce. However, this relationship varied across participants in analytically significant ways.

\subsection{Perception: how epistemic and ethical concerns become intelligible}

Participants invoked similar ethical concepts, including privacy, harm and accountability, but understood and interpreted them differently. These differences revealed distinct but overlapping moral economies: patterned orientations towards what counts as valuable knowledge, legitimate intervention and responsible practice. Individuals drew on several, and those in similar settings could articulate different concerns.

\emph{Moral economy of exposure, agency and safety:} In this moral economy register, the affected person's safety, dignity, agency and control over information became especially salient. For N, whose organisation developed a chatbot providing information and support to potentially exploited individuals, collecting and retaining personal information was understood through the risk of exposing users to harm. N referred to circumstances in which a person might need to leave an interaction quickly \emph{``if you're being monitored.''} Information that could support personalisation or continuity of assistance could also increase users' vulnerability through traceability, so the ethical concern for safety shaped what was appropriate to know. G, a researcher working with trafficking-related data in an international organisation, linked ethical concerns to institutional responsibilities for protecting affected populations. As G explained\emph{, ``I specifically am attached to the protection part, which of course means that we have certain concerns when it comes to data...''} G sought to enable the analysis and sharing of victim-assistance data while mitigating re-identification risks through synthetic data. F asked ``\emph{who's got the technology,}'' observing that smaller NGOs and safe houses lack comparable funding and technical capabilities despite having direct access to survivors' sensitive information. F described organisations sharing spreadsheets by email without adequate safeguards because ``\emph{they didn't even have that knowledge and have that capability.}'' F warned that using ChatGPT for survivor-related questions could similarly mean ``\emph{giving the data of the survivor away to such LLM}.'' Privacy appeared as a mismatch between responsibility for survivors and capacity to control the technologies processing their information. B, a researcher involved in developing a worker-facing screening tool, recognised that ``\emph{we may have technical expertise \ldots{} but we are not experts of domains}'', and treated engagement with ``\emph{real users who we could meet with}'' as able to \emph{``quickly change the direction we were going.''} Recognising the limits of developers' knowledge created an epistemic rationale for participation, while respecting workers' agency provided an ethical orientation for technological development.

\emph{Moral economy of hidden exploitation, evidence and actionability:} In investigative and intelligence-oriented accounts, the dominant ethical concern was failing to identify hidden exploitation, which oriented epistemic priorities towards linking fragmented information and generating leads, quite different from the protective rationale. A was involved in data-driven anti-trafficking investigations, and described trafficking becoming visible through patterns and anomalies across a range of public and administrative data (i.e., visa categories, travel histories, addresses and financial records), seeking investigative cues. Where protection-oriented accounts valued compartmentalisation and affected-person control, A derived epistemic value from aggregation, linkage and scale \citep{aradau2022algorithmic}. The same orientation that endorsed aggregation also shaped which harms were perceptible. Asked whether demographic indicators could contribute to nationality-based or racialised profiling, A said: \emph{``I don't know how much more to add to that, sorry''.} This potential ethical concern is neither explicitly rejected nor actively defended. Instead, it remains difficult to articulate, suggesting how moral economies may shape the boundaries of ethical perception: concerns aligned with institutional purposes become readily recognised, while others may remain less salient or lack an established frame through which they can be addressed. For E, who works with dark-web intelligence and open-source investigations, data were an investigative resource: ``\emph{Netflow data, Adtech data, OSINT data and also closed sources},'' subjected to ``\emph{data point exploitation}.'' E's statement was direct: \emph{``I only care about GDPR as I protect the data that I have. Other than that, if people are being exploited, there's human rights violations, then they have no right to privacy. I'm going to exploit those data points as I see fit, regardless of any repercussions.''} Privacy was narrowly framed around the lawful and compliant handling of data under E's control. For N, traceability is a route to exposure and potential harms; for E, traceability is a route to detection. K's account complicates this register. For them, operational usefulness remained important, but they also questioned whether the data and categories through which patterns became intelligible were themselves valid and ethically defensible. While building tools for police and border settings, K warned of \emph{``biasing the training datasets''} when agencies procure systems without understanding how those systems were developed, drew attention to the ``\emph{grey area between trafficking, smuggling, illegal immigration}'', and problematised the surveillance implications that might follow---concerns less salient for A, who treated biased data as \emph{``noise'',} or for E. K's academic formation and engagement across professional settings may have exposed them to multiple interpretive repertoires, suggesting that perception might be shaped by professional formation as well as institutional purpose.

\emph{Moral economy of risk, commercial interests and compliance:} Where the object of analysis was companies and supply chains rather than individuals experiencing exploitation, concerns about trafficking and exploitation were mediated through organisational risk, Environmental, Social and Governance (ESG) frameworks, regulatory obligations, investor expectations and reputational exposure. H, whose work focuses on enterprise technology and ESG systems, described a platform that classifies supply chains as \emph{``low risk, high risk, medium risk ... based on external data,''} translating dispersed information into categories of organisational risk, rather than investigative cues to track. This classification involved epistemic choices about which indicators constitute trafficking risks, with ethical implications for how suppliers and workers are represented and potentially affected by organisational decisions. C, a researcher analysing corporate modern slavery statements, noted that making corporate reporting practices computationally visible does not reveal actual conditions of exploitation or establish corporate accountability and was \emph{``so far removed from being involved with survivors.''} Worker harm became most salient when translated into organisational exposure, while workers' experiences and claims for remedy remained less represented. The pattern was not confined to firms: advising civil society organisations, participant I observed that \emph{``the customer of a non-profit project is the donor. It's not the beneficiary''}.

\emph{Moral economy of epistemic authority, validity and representation:} Some participants located ethical concern in the wider ecosystem. Responsibility in this register directs concerns to the relationship between representation, knowledge claims and the social phenomenon being represented. F highlighted how epistemic and ethical concerns could become diluted as knowledge passed across professional and organisational boundaries, observing that developers \emph{``really only know the technical side of things,''} while domain knowledge and requirements often pass through \emph{``a lot of layers in between}'' NGOs, technology providers, managers and developers. Discussing the need for coordination across digital platforms, M observed that \emph{``each of these types of platforms or actors see the issue from a different angle.''} F further questioned \emph{``how could you know if a tool is effective if the survivor voice and perspective was not taken into account?''}, and criticised register-neutral ethical vocabulary failing at the field level: actors \emph{``just speak words like accountability---exactly what are you talking about? It's like the word technology that they use to get funding for an anti-human trafficking project.}'' O questioned whether applying Western understandings of human rights in Asia risked becoming ``\emph{another form of colonialism\ldots{} using sort of a Western value system, laying it over an Asian value system}'', and advocated developing a shared ethical vocabulary through multicultural dialogue, to \emph{``define terms which are not just from one regional understanding and imagination, but actually are multicultural.''}

Concerns were also epistemological. G cautioned that changes in recorded cases were \emph{``not actually a representation of\ldots{} how many people were exploited; it's simply a change in how many people were assisted.''} O distinguished \emph{``hard data''} from \emph{``interpreted data''} in the context of forced labour supply chain detection, arguing the need for getting information directly from workers \emph{``that's why we've gone for the workers, because we're thinking it's much better hard data that you get.''} D, developing a prototype for assessing forced labour risk in the maritime sector using vessel and administrative data, observed that: \emph{``when we see people as data, we're abstracting out lots of things}'', losing \emph{``a lot of the context and what's actually happening''} and potentially depriving people of \emph{``their own autonomy or agency.''} Similarly, F warned increased use of AI could push \emph{``the human suffering, human emotion, human side even further to the background.''} Here, representation became ethically problematic when communities were made objects of observation rather than participants in producing knowledge.

\subsection{Judgement: how concerns are evaluated, prioritised or dismissed}

Concerns perceived within these four moral economies were not, by themselves, actionable. Participants described situated judgements that involved weighing concerns against other ethical considerations, such as privacy versus detection; against institutional imperatives such as operational effectiveness, feasibility, and funding; and against epistemic questions about sufficient evidence and acceptable uncertainty. Moral economies shaped which concerns carried weight, while institutional logics influenced what was considered legitimate or feasible, together determining which epistemic and ethical concerns were prioritised, justified or subordinated.

E argued that \emph{``protecting that data at all costs is kind of a hindrance to the cause''} and justified extensive aggregation by the seriousness of exploitation when referring to child trafficking investigations. K, also working with law enforcement, acknowledged the benefits of linking personal data but judged that moving to individual identification made ``the risk of the system\ldots much higher'', ``a completely different category'', and considered the resulting limits on investigative capability \emph{``completely justified for human rights.''}

In commercial settings, H was explicit that ``\emph{most of the companies don't care about it\ldots{} it's always investor pressure when you're a large public company,''} and \emph{``the reality is we still need to drive revenue first.''} The weight of worker harm depends on its connection with relevant institutional logics, in this case, of business incentives, legal exposure, or reputational risk. Participant I was sceptical about whether corporate risk reduction helped improve conditions for workers: \emph{``there's this Venn diagram between tackling corporate risk and tackling worker risk\ldots If you're just dealing with finding risks as a corporation \ldots{} just avoiding those risks, then to a worker, actually that doesn't feel particularly conducive to attacking forced labour either.''}

In other cases, anticipated harms would constrain action. B considered whether a worker-facing tool should collect photographs to distinguish workers with similar names. Although frontline practitioners were in support, the development team weighed \emph{``the value versus the risk of harm''} and concluded that photographs presented \emph{``too much risk of harm''} and were \emph{``not in the worker's best interest.''} Here, judgements about the legitimate epistemic function of the tool were inseparable from concerns about worker safety.

Judgements were also made about how to make judgements under conditions of uncertainty. N described ongoing deliberations over a \emph{``warm handoff''} feature connecting chatbot users with human advocates that raised concerns about consent, privacy and the safe transfer of sensitive information. Against the prevailing industry disposition to \emph{``move fast and break things and fix them later,''} N's judgement was explicit: \emph{``we don't believe in that.''}

\subsection{Situated action: enacting, constraining and displacing judgement}

Whether judgement became action depended on authority, resources, professional responsibilities and organisational priorities. Participants described a range of strategies.

\emph{Stretching institutional practices}. Participants sought to introduce alternative epistemic and ethical priorities to their institutional settings. Within a security-oriented setting, K sought to bring their academic background to bear on institutional practice and proposed ``\emph{a stronger humanitarian dimension to the security processing''}, including questioning conventional vulnerability classifications, \emph{``if you prioritise only women and children, it means that men can never have high vulnerability.''} O sought to extend corporate supply-chain monitoring beyond conventional auditing approaches by incorporating worker-generated knowledge: ``\emph{Who knows more about what's happening than anybody else? OK, the workers do.''} O described building a digital platform that collected information directly from workers and compensated them for their contributions. Both drew on a moral economy at odds with the dominant logic of their setting to introduce alternative epistemic priorities.

\emph{Negotiating boundaries}. G kept victim-assistance information separate from investigation---``we don't share that information with law enforcement''---and, to enable research while protecting sensitive data, used synthetic data and differential privacy against re-identification from successive releases. G acknowledged that this might not hold: reverse engineering was not possible ``\emph{at least not currently. It might be in the future\ldots we'll deal with that when we cross that''.} Negotiation here meant establishing enough confidence in safeguards to proceed while deferring an acknowledged future harm. H described longstanding involvement in charitable activities, while recognising that in their commercial role, they also had \emph{``the capability professionally to probably do something there.''} H sought to reconcile social objectives with business viability: \emph{``how do you drive a lot of revenue faster and then start flowing it into it.''} Their activities remained bounded by commercial logic, with social objectives pursued through, rather than independently of, revenue generation.

\emph{Reproducing institutional practices}. Some actors responded to epistemic and ethical concerns through professional repertoires congruent with the institutional logics within which they worked. A recognised the \emph{``dark figure''} of trafficking, observing that some victims \emph{``don't come forward''} while others are \emph{``deported before they're even investigated,''} arguing that \emph{``these sorts of indicators should be put into the prevention focus.''} An investigative orientation shaped which solutions appear appropriate and legitimate to A: missing knowledge was interpreted as a detection problem, making more indicators a plausible response within established professional practices.

There were two ways in which action was limited. On the one hand, some participants reached an ethical judgement but lacked the institutional conditions required to enact it (\textbf{constrained action}). D concluded that responsible development of a forced-labour risk tool required sustained engagement with those expected to use or be affected by it: ``\emph{If I had the resources, I would embed myself into whatever the community or the organisation who is supposed to use the tool}.'' Yet the fellowship provided neither resources, connections nor mandate. D subsequently described losing motivation. C reflected, ``\emph{I wish we would have found better ways in which we could have engaged survivors into the process}.''

On the other hand, some participants \textbf{displaced responsibility} elsewhere. L, a technology developer, argued that ethics belongs to the ``applicator'' of the technology rather than the builder, arguing that \emph{``the application of the technology is always unknown from the creator's perspective''} and that \emph{``it's always on the applicator of that technology to question the ethics of how they're using it.''} E pointed to the end of the chain: after \emph{``multiple layers of validation\ldots{} this is up to the courts to decide''.} In neither case is the concern unperceived; it is perceived and then routed away, to other actors in the AI development chain, despite the importance of upstream epistemic choices embedded in technical artefacts, shaping what downstream actors can know, interpret and act upon \citep{akrich1992de,aradau2022algorithmic,suchman2007human}.

Across these accounts, we found that situated agency operated through, rather than independently of, institutional logics. Importantly, ethical recognition did not guarantee action, and action itself did not necessarily resolve the underlying epistemic and ethical concerns.

\subsection{Socio-technical materialisation: what becomes embedded and sustained}

Socio-technical materialisation concerns how these responses become built (or not built) in the systems, workflows, data architectures, or organisational routines; which ones were narrowed, displaced, or left unsupported---and what happened when artefacts travelled across time and space.

Some commitments were materialised through concrete design features. N designed on the assumption that \emph{``somebody might be using it with an abuser looking over their shoulder.''} This understanding of vulnerability informed several features: no accounts, no IP collection, sessions that reset after two hours of inactivity, a quick-exit button and contractual limits on third-party retention. B's tool excluded photographs, restricted access to voice recordings, and asked workers whether they wanted help leaving their situation, so that worker agency sat alongside, rather than being inferred from computational indicators.

E translated commitments to responsible AI into procedures---\emph{``label it, audit it, validate it''} through spot checks for bias and audits of data provenance. These are meaningful epistemic safeguards; however, responsibility is operationalised as quality control that narrows the concern to the reliability of outputs rather than the legitimacy of classification or data collection.

Other commitments informed artefacts without reaching practice. C's interpretable model for analysing modern slavery statements was not yet deployed; K's protection-oriented vulnerability model saw limited uptake; D's prototype remained unfinished. Even deployed tools depended on arrangements beyond the project: F described tools that \emph{``are not being continued''} after staff left. B's tool survived a change of institution as a partner organisation took responsibility. As it expanded internationally, B's team adapted questions, languages, interfaces, and assessment frameworks---\emph{``the system itself is designed to be adapted like that''}---which raises the question of which original commitments remain stable and whose knowledge informs later modifications.

\subsection{Reflexivity and the conditions of situated agency}

This section concerns how actors could critically engage with the institutional frameworks shaping their practices and what enabled or constrained their capacity to act differently. An important distinction emerged between reflection \emph{within} professional frameworks and reflection \emph{on} those frameworks. \emph{Within} investigative frameworks, E, for example, was concerned with the reliability of investigative information through validation, auditing and data quality. On the other hand, other participants questioned whether their existing knowledge, categories or understandings of responsibility were themselves sufficient. H acknowledged having \emph{``a lot of blind spots''} from a career in commercial technology with limited direct experience of affected communities. F, reflecting on earlier civil-society work after acquiring greater technical expertise, said \emph{``you didn't know what you didn't know, so you did your best.''} K challenged conventional vulnerability categories while O questioned the assumptions of established human-rights approaches. L extended responsibility in time---``\emph{How will it live long enough beyond your own participation to see whether it will do good or harm?''}---wishing ``\emph{to see it through in a way that I think is more ethical for my own morality,''} and in hindsight, to \emph{``find a good customer and put those tools in a way that they sustained beyond the program.''} G reflected on their position in a small research unit within an implementation-oriented organisation, where \emph{``the step from research to practical implementation is quite short'',} yet colleagues \emph{``might not always see\ldots the added value of what we do''.}

Throughout the accounts, whether recognised concerns could be acted upon depended on at least three conditions\emph{. First,} discretionary time and resources. B attributed the team's fieldwork and iteration to the ``luxury of time because we are in a research institute''; D and C, as shown above, lacked such slack. \emph{Second,} exposure to more than one register. Reflexivity was particularly visible among actors formed across settings---K between academia and law enforcement; F between civil society and development; B between computer science and field work---and among those with sustained contact with affected people, as in B's ``real users'', and N's co-creation with people with lived experience. \emph{Third}, mandate and organisational continuity. G could refuse to share case data as their unit was ``attached to the protection part'', a mandate that legitimised a refusal other actors might contest; and whether decisions outlived a project depended on organisational relationships, as F's discontinued tools and B's transferred tool show.

\section{Implications and conclusions}

This section considers the implications of the three issues identified in the Introduction for responsible AI.

\subsection{Beyond a single principle--practice gap}

The study has shown that the principle--practice gap is more complex than a linear problem of implementation. Principles and practices are recursive, echoing \citet{ruster2025gaps}, and the findings point to different analytical moments at which epistemic and ethical concerns can be lost or weakened. As Table~\ref{tab:attrition} shows, concerns could go unperceived, be outweighed, be blocked, be displaced, or be embedded and then narrowed, left undeployed or not sustained. Each point of loss requires different governance instruments, beyond ethics training and principles.

\begin{table}[t]
\caption{Points of attrition of epistemic and ethical concerns and governance implications.}
\label{tab:attrition}
\Description{A four-row table. Rows are the analytical moments perception, judgement, situated action and socio-technical materialisation. Columns describe how concerns are lost or weakened at that moment, the participant accounts that illustrate it, and the corresponding governance implications.}
\footnotesize
\begin{tabular}{@{}p{0.13\linewidth}p{0.27\linewidth}p{0.18\linewidth}p{0.35\linewidth}@{}}
\toprule
\raggedright \textbf{Analytical moment} & \raggedright \textbf{How concerns are lost or weakened} & \raggedright \textbf{Illustrative accounts} & \raggedright \textbf{Governance implications} \tabularnewline
\midrule
\raggedright Perception & \raggedright Concerns remain unrecognised or insufficiently articulated within the actor's professional or institutional understanding. & \raggedright A --- limited recognition of profiling harms & \raggedright Bring alternative forms of knowledge and institutional perspectives into problem formulation, including affected-community perspectives; create opportunities to question prevailing assumptions. \tabularnewline
\raggedright Judgement & \raggedright Concerns are recognised but outweighed by competing priorities, without clear justification of the trade-offs. & \raggedright E --- privacy versus detection; H --- worker harm versus commercial objectives & \raggedright Make trade-offs explicit and open to scrutiny: what concerns were considered, whose interests were prioritised, who made the decision and on what grounds. \tabularnewline
\raggedright Situated action & \raggedright Concerns are judged important but actors lack the resources, authority or mandate to act, or responsibility is displaced to other actors. & \raggedright C, D --- institutional constraints; E, L --- displaced responsibility & \raggedright Provide resources, decision-making authority and institutional support; clarify responsibilities across development, deployment and downstream use. \tabularnewline
\raggedright Socio-technical materialisation & \raggedright Concerns inform development but are only partially incorporated, narrowed, left undeployed or not sustained over time. & \raggedright E, C, K, D, F & \raggedright Trace how ethical judgements are translated into design and deployment; review what was incorporated, modified or excluded; establish responsibility, resources and continuity for implementation and maintenance. \tabularnewline
\bottomrule
\end{tabular}
\end{table}

\subsection{Epistemic choices as sites of ethical significance}

The analysis has shown how epistemic and ethical considerations are intertwined in practice. Participants made varied epistemic choices, sometimes with and sometimes without recognising their ethical significance. Some decisions were approached and understood as technical matters (e.g., the construction of a forced-labour risk score), without explicitly foregrounding how choices of indicators, categories and inference might have ethical consequences themselves. In other cases, participants deliberately used epistemic solutions as a means to manage ethical concerns, inscribed in the systems (e.g., not linking records, not collecting photographs). These choices, as shown, were varied, emerged through practice, and were negotiated and acted upon under conditions of uncertainty. A procedural ethical guideline or checklist alone may therefore be too rigid to accommodate this indeterminate and dynamic process. At the same time, separating technical validation from ethical review would risk overlooking how these concerns inform one another, and leaves non-inscription---what was deliberately not collected, linked or inferred---invisible to assessment. A more integrated approach would examine how these epistemic and ethical choices are made and negotiated at each stage of AI development. By documenting the reasoning behind development decisions---assumptions considered, concerns raised, trade-offs negotiated, and alternatives rejected---epistemic choices are open for scrutiny and, in effect, support meaningful human oversight from the design stage. This resonates with \citet{mitchell2026ai}, who argued that ``humans-in-the-loop'' at the end stage as users do not ensure effective oversight if users lack the knowledge, cognitive capacity or practical control to critically evaluate and intervene in AI-generated decisions. 

\subsection{Building epistemic infrastructure}

Participants affirmed the importance of survivors' experiences, needs and agency, yet few accounts described direct engagement with affected populations. Building on STS understandings of infrastructure as relational and embedded in practice \citep{star1994steps}, we suggest strengthening epistemic infrastructure for responsible AI: the \emph{durable} institutional arrangements through which different forms of knowledge can inform technological development; epistemic and ethical concerns can be recognised and evaluated; and the resulting judgements can meaningfully influence technological decisions. In practice, this goes beyond one-off stakeholder consultation or collaboration. It involves, for example, dedicated resources for sustained engagement across technical, domain and affected-community perspectives; joint processes for examining problem definitions, data and modelling assumptions; and mechanisms through which epistemic and ethical concerns can inform or challenge development decisions. Access to additional knowledge alone risks becoming another form of \emph{participation washing} or \emph{ethics washing;} what matters is whether that knowledge is reflected in decisions. At field level, funders, technology providers and implementing organisations can support this through resource allocation, shared responsibilities and requirements to document how competing forms of knowledge inform development choices.

In conclusion, this study shows that Responsible AI is understood and enacted through situated practices that are both epistemic and ethical. Different institutional logics and moral economies shaped what participants recognised as responsible and how these judgements were translated into design and development decisions. Therefore, improving Responsible AI requires attention to the institutional conditions through which knowledge is produced, contested and allowed to influence technical choices. Future research could extend these findings through ethnographic, participatory and longitudinal studies with affected communities.

\section*{Funding}
This work was supported by the Engineering and Physical Sciences Research Council (EPSRC) (grant EP/S023437/1).

\section*{Ethical considerations}
The study received ethics approval from the University of Bath before data collection (reference 6635-10859). All participants took part voluntarily and gave informed consent, including consent for audio recording and transcription, after receiving information about the study's purpose, the use of their data and their right to withdraw. Interviews concerned professional practice; participants were not asked about their own experiences of exploitation, and no survivors or affected people were interviewed. Because the field is small and specialised, organisational affiliations, job titles, named technologies and locations are reported at a general level, and participants are identified by letter codes to reduce re-identification risk. Information about safeguarding and support services was made available.

\section*{Generative AI usage statement}
During the preparation of this work, the authors used ChatGPT (OpenAI; including GPT-5.6), Claude (Anthropic) and Microsoft Copilot to support language editing, organisation and formatting of manuscript text. GenAI tools were not used for interview data analysis. The authors retained responsibility for and independently developed all conceptual, methodological, analytical and interpretive decisions as well as the final manuscript.

\bibliographystyle{plainnat}
\bibliography{references}

\clearpage
\appendix
\section{Participants}\label{app:participants}

The table summarises the 14 participants interviewed between October 2025 and February 2026. To reduce the risk of re-identification in a small, specialised field, organisational affiliations, job titles, named technologies and locations are reported at a general level, and participants are identified by letter codes.

\begin{table}[H]
\caption{Participant roles, institutional settings and applications}
\label{tab:participants}
\Description{A table listing the 14 interview participants by letter code, with their institutional setting, professional role and the type of AI application discussed.}
\footnotesize
\begin{tabular}{@{}p{0.06\linewidth}p{0.25\linewidth}p{0.30\linewidth}p{0.32\linewidth}@{}}
\toprule
\raggedright \textbf{Code} & \raggedright \textbf{Institutional setting} & \raggedright \textbf{Professional role} & \raggedright \textbf{Type of application discussed} \tabularnewline
\midrule
\raggedright \textbf{A} & \raggedright Law enforcement/start-up & \raggedright Practitioner with law-enforcement and investigative experience & \raggedright Data-driven investigation linking administrative and financial records \tabularnewline
\raggedright \textbf{B} & \raggedright Research institute & \raggedright Researcher (computing) working with frontline organisations & \raggedright Worker-facing screening and support tool \tabularnewline
\raggedright \textbf{C} & \raggedright Academia & \raggedright Researcher & \raggedright Automated analysis of corporate modern slavery reporting \tabularnewline
\raggedright \textbf{D} & \raggedright Academia & \raggedright Researcher & \raggedright Forced-labour risk assessment in a maritime setting \tabularnewline
\raggedright \textbf{E} & \raggedright Private sector & \raggedright Investigative technology developer & \raggedright Online and open-source intelligence analysis \tabularnewline
\raggedright \textbf{F} & \raggedright Civil society and technology development & \raggedright Data and AI developer with prior NGO experience & \raggedright Data tools for NGOs and survivor-support organisations \tabularnewline
\raggedright \textbf{G} & \raggedright International organisation & \raggedright Researcher in a research and evidence unit & \raggedright Privacy-preserving sharing of victim-assistance data \tabularnewline
\raggedright \textbf{H} & \raggedright Private sector & \raggedright Enterprise technology and ESG lead & \raggedright Supply-chain risk assessment platform \tabularnewline
\raggedright \textbf{I} & \raggedright Civil society and advisory & \raggedright Adviser across NGOs, worker voice and supply-chain governance & \raggedright Worker-voice and supply-chain initiatives \tabularnewline
\raggedright \textbf{K} & \raggedright Academia, with law-enforcement and border partners & \raggedright Researcher & \raggedright Analytical tools and vulnerability assessment for police and border settings \tabularnewline
\raggedright \textbf{L} & \raggedright Technology sector & \raggedright Software engineer & \raggedright Tools for collecting and analysing online data for investigations \tabularnewline
\raggedright \textbf{M} & \raggedright Cross-sector technology partnership & \raggedright Programme and partnership lead & \raggedright Technology support programme for anti-trafficking organisations \tabularnewline
\raggedright \textbf{N} & \raggedright Non-profit, survivor support & \raggedright Programme and partnership lead with experience in survivor services & \raggedright Conversational support tool for potentially exploited people \tabularnewline
\raggedright \textbf{O} & \raggedright Civil society and cross-sector partnerships & \raggedright Programme lead & \raggedright Digital worker reporting and supply chain monitoring platforms \tabularnewline
\bottomrule
\end{tabular}
\end{table}

\end{document}